# It's Time to Consider 'Time' when Evaluating Recommender-System Algorithms [Proposal]

Joeran Beel
Trinity College Dublin
Department of Computer Science and Statistics
Knowledge and Data Engineering Group
ADAPT Centre, Ireland
joeran.beel@adaptcentre.ie

## ABSTRACT

In this position paper, we question the current practice of calculating evaluation metrics for recommender systems as single numbers (e.g. precision *p=.28* or mean absolute error *MAE = 1.21*). We argue that single numbers express only average effectiveness over usually a rather long period (e.g. a year or even longer). This practice provides only a vague and static view of the data. We propose that recommender-system researchers should instead calculate metrics for time-series such as weeks or months, and plot the results in e.g. a line chart. This way, results show how algorithms' effectiveness develops over time, and hence the results allow drawing more meaningful conclusions about how an algorithm will perform in the future. In this paper, we explain our reasoning, provide an example to illustrate our reasoning and present suggestions for what the community should do next.

## 1  INTRODUCTION

Recommender-system evaluation is an actively discussed topic in the recommender-system community. Discussions include advantages and disadvantages of evaluation methods such as online evaluations, offline evaluations, and user studies [1–4]; the ideal metrics to measure recommendation effectiveness[1] [5–8]; and ensuring reproducibility [9–11]. Over the last years, several workshops about recommender-system evaluation were held and journals published several special issues [11–14].

An issue that has received (too) little attention is the question if presenting results as a single number is sufficient or if metrics should be presented for time intervals. Typically, researchers calculate a few metrics for each algorithm (e.g. precision **p**, normalized discounted cumulative gain **nDCG**, root mean square error **RMSE**, mean absolute error **MAE**, coverage **c**, or serendipity **s**). For each metric, they present a single number such as *p = 0.38, MAE = 1.02*, or *c = 97%*, i.e. the metrics are calculated based on all data available. Hence, the metrics express how well an algorithm performed on average over the period of data collection, which is often rather long. For instance, the data in the MovieLens 20m dataset was collected over ten years [15]. This means, when a researcher reports that an algorithm has e.g. an RMSE of 0.82 on the MovieLens dataset, the algorithm had that RMSE *on average* over ten years.

## 2  THE PROBLEM: SINGLE-NUMBER METRICS

We argue that presenting a single number that expresses the overall average is problematic as an average provides only a broad and static view of the data. If someone was asked how an algorithm had performed over time – i.e. before, during, and after the data collection period, the best guess, based on a single number, would be that the algorithm had the same effectiveness all the time.

Consider the following example: A researcher aims to compare the effectiveness of algorithms *A* and *B*. She receives 12-months-usage data from a recommender system in which the two algorithms have been used [2]. The researcher calculates for algorithms *A* and *B* a metric (e.g. precision) to express the algorithms' effectiveness (a real researcher would probably calculate more than one metric but to illustrate our point, one metric is sufficient). The outcome of such an evaluation would typically be a chart as presented in Figure 1. The chart shows the effectiveness for algorithm A (0.48) and for algorithm B (0.67). The interpretation of these results would be that algorithm B is more effective than algorithm A, and algorithm B should be used in a recommender system rather than algorithm A.

---

[1] We use the term 'effectiveness' to describe to what extent a recommender system achieves its objective, which could be, for example, maximizing user satisfaction (measured e.g. through user ratings) or revenue. However, for this paper, it is not important what the actual objective of the recommender system is or which metric is used.

[2] Please note that for our argument, it would not matter if data is used from a real-world recommender system that implements the algorithms (as in our example), or if a researcher uses a dataset like the MovieLens dataset.

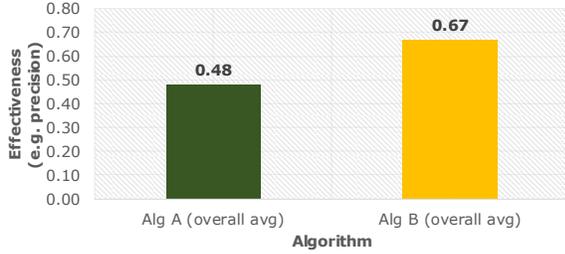

*Figure 1: Results from a hypothetical, yet typical, evaluation of two recommendation algorithms A and B*

If someone was asked how algorithms A and B had performed in the past, and will perform in the future, the best guess would be that the algorithms' effectiveness was stable and will remain stable over time. This assumption is illustrated in Figure 2.

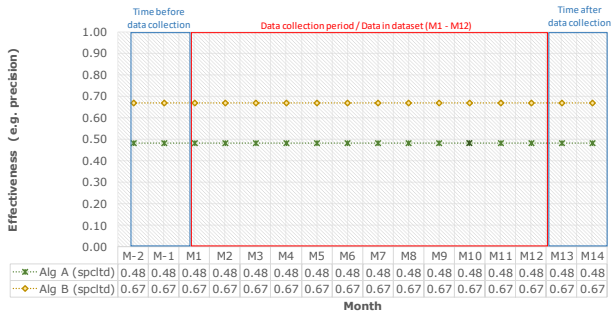

*Figure 2: Effectiveness of algorithms A and B over time. The numbers are best-guesses based on the results in Figure 1. The time during which data was collected are months M1 to M12, the time before the data collection period are months M-1 and M-2, and the time after the data collection period are months M13 and M14.*

We argue that such assumptions are naïve, as many algorithms' effectiveness is not stable over time. It is well known that the effectiveness of many recommendation algorithms depends on the number of users, items, and ratings as well as algorithm's parameters such as neighbourhood size or user model size [16–19]. As the numbers of users etc. are likely to change over time, also the effectiveness of the algorithms will change over time. We have observed this effect in our own recommender systems Docear [20] and Mr. DLib [21], as have Middleton, Shadbolt, & De Roure [22] and Jack from Mendeley [23]. For instance, Jack reports that precision increased from 0.025 when Mendeley launched its recommender system to 0.4 after six months. Also Netflix emphasizes the importance of considering time in recommender systems [24].

## 3 OUR PROPOSAL: TEMPORAL EVALUATION

We propose that, instead of a single number, recommender-systems researchers should present metrics for time series, i.e. each metric should be calculated for a certain interval of the data collection period, e.g. for every day, week, or month. This will allow to gain more information about an algorithm's effectiveness over time, identify trends, make better predictions on how an algorithm will perform in the future, and hence to make more meaningful conclusions on which algorithms to deploy in a recommender system.

Calculating effectiveness for each month would lead to a chart like in Figure 3, given the data from the previously introduced example. The chart shows that effectiveness of algorithm A improves over time from 0.14 in the first month of the data collection period to 0.90 in the last month. In contrast, the effectiveness of algorithm B decreases from 0.83 to 0.53. Most importantly, the chart shows that algorithm A outperformed algorithm B from month nine onwards.

Even though Figure 1 and Figure 3 are based on the same (hypothetical) data, Figure 3 is more meaningful than Figure 1. Based on Figure 1, one would conclude that algorithm B is more effective than algorithm A. Based on Figure 3, a more differentiated conclusion can be drawn, namely that algorithm B was only more effective during the first months, but after month 9, algorithm A was more effective, and looking at the trend it seems likely that algorithm A will continue to be the more effective algorithm in the future.

We therefore propose that recommender-system researchers should calculate their metrics for time-intervals of the data collection period, and present them in line plots as shown in Figure 3. Similarly, reviewers and organizers of conferences and journals should encourage the submitting authors to present their results as time series, when possible. Also, researchers who publish datasets should include time information such as when a user registered or when a rating was made.

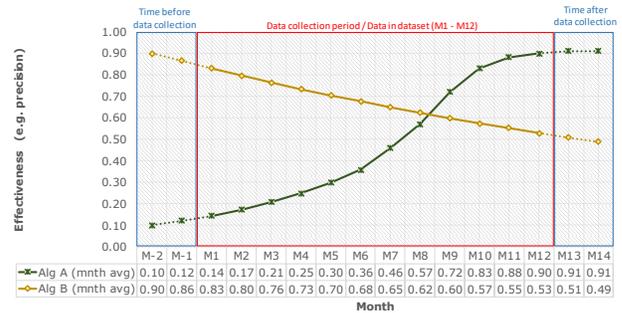

*Figure 3: Hypothetical results from the same evaluation as in Figure 1, but plotted over time with monthly averages. Months M1-M12 are based on the available data. Months M-1, M-2, M13, and M14 are predicted.*

It is a common practice already to analyze how algorithms react to changes in the algorithms' parameters or the number of items and users in a dataset [16], [19]. Researchers analyze, for instance, how effective an algorithm is with a neighborhood size of two, three, four, etc. or how effectiveness changes based on the number of data points in a dataset. While these information certainly is relevant, no one currently knows, how many and which variables affect an algorithm's effectiveness [10]. Therefore, it is not possible to present a comprehensive analysis of all variables effecting an algorithm's effectiveness. We consider 'time' to be a good aggregate, and we think that knowing how an algorithm's effectiveness changes over time is at least equally important as knowing how it changes based on variations in e.g. neighborhood size, user model size or the number of users.

## 4 RELATED WORK

"Time" has been considered in recommender-system research, though usually not for evaluations. The user-modelling community considers time in terms of concept drift, i.e. changes in user preferences over time [25–32]. Their focus lies on adjusting algorithms to make them consider time and improve the effectiveness of the algorithms. "Time" has also been considered as contextual feature, i.e. depending on the temporal context (e.g. *summer* or *winter*), different recommendations should be given [33–36], or different algorithms should be used [10]. There is also work on enhancing recommendation algorithms by considering temporal data, mostly in the field of collaborative filtering [37–41]. Again, this work focuses on incorporating temporal data into the algorithms. The machine-learning community sometimes considers how temporal aspects can be taken into account for training and testing algorithms, and how predictions for time-series can be made [33], [42], [43]. For instance, typical k-fold cross validations may be adjusted to not randomly pick training and test data, but train algorithms only on data from a certain period, and evaluate the algorithm on data from the subsequent period (illustrated in Figure 4). Similarly, re-training machine-learning systems after certain periods is a common research topic [44].

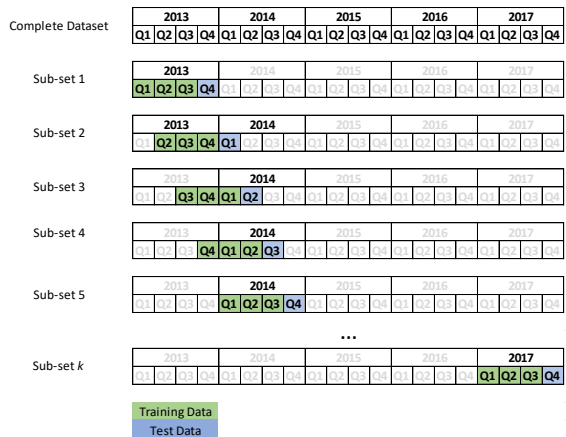

*Figure 4: Illustration of time-aware machine-learning training and testing.*

To the best of our knowledge, there has been little work on how to calculate and present evaluation metrics over certain time periods. In addition, even though there has been some work, the majority of the recommender system community seems not yet to consider it, i.e. it continues to present single number results. Rana & Jain developed a time-aware book recommender system and evaluated it per week and quarter [45]. However, they did not evaluate the performance over time but the preferences of the users, i.e. if users preferred rather diverse or similar recommendations (the users´ preferences changed over time). Lathia et al. showed that a "most popular" recommendation algorithm on the MovieLens dataset converges towards a random baseline over time. Soto evaluated several algorithms with different metrics on the Movielens 1m dataset and plotted the effectiveness of the algorithms over a period of 20 months [46]. The results show that effectiveness of the algorithms often varies over time. However, only in few cases, the variations were so strong, that a conclusion about which algorithm was most effective would have changed compared to looking at single number results.

## 5 NEXT STEPS

We used a hypothetical example to demonstrate the need for time-based evaluation metrics. The related work shows a few examples in which algorithms´ performance varies over time, but the variations were usually not very strong. Consequently, we first need to identify if and to what extent the need for evaluating recommender systems over time really exists. We suggest to analyze existing datasets such as MovieLens [47], RARD [47], Docear [20] or other datasets. The analyses should calculate metrics for different algorithms over time, to see if and how strong the effectiveness of algorithms changes. It will be particularly interesting to see how often the change is so strong that the conclusions about which of two algorithms is more effective will change. To further quantify the problem, a literature survey could be conducted to find out how many researchers currently present single-number metrics, and how often time-series metrics might make sense. A suitable corpus to analyze would be the full papers from the previous ACM Recommender Systems conferences (see appendix).

If the research confirms our assumptions, specific time-series metrics need to be created. One option would be, as done in the example, to calculate each metric e.g. per month and plot the results in a chart. However, in some cases, space restrictions might prevent researchers from presenting numbers for each interval. In such cases, it might be sensible to present the min, max, and average values for the intervals as well as standard deviation; or the values for the first and last month and/or a trend function. The community should also agree on notations for the time-series metrics. For instance, to express precision $p$ in interval $i$, the metric could be labelled $p@i$ (e.g. p@m5 to express average precision in the fifth month).

## FILE HISTORY

- 2017-10-29: Removed the list of RecSys publications from the appendix. Added a related work section (Acknowledgements to Michael Ekstrand for pointing to some references).
- 2017-09-02: Added RMSE as metric; some minor changes (typos, grammar, …).
- 2017-08-30: Moved list of RecSys papers to appendix; Added reference to Netflix paper that mentions the importance of time; adjusted the header of the document.
- 2017-08-28: First version